\documentstyle[preprint,aps]{revtex}
\begin{document} 
\tightenlines
\draft

\title{ On the nuclear states of $K^{-}$  mesons  }
\author{S.~Wycech\thanks{email:  wycech@fuw.edu.pl}} 
\address{Soltan Institute for Nuclear Studies, Warsaw, Poland}
\author{A.M.~Green\thanks{email: anthony.green@helsinki.fi}}
\address{Helsinki Institute
of Physics, P.O. Box 64, FIN--00014 University of Helsinki, Finland}

\date{\today}

\maketitle

\begin{abstract}
 The $ KNNN$ bound states  recently discovered at KEK   are studied.
It is shown that the $\Lambda(1405)$ and $ \Sigma(1385)$ resonant   
states  coupled to  the $KN$ system may generate an  attraction strong 
enough   to form  such  bound  states.  
\end{abstract}

\section{Introduction}
\label{intro}

The  bound   $KNNN$  states,   recently discovered at KEK,  
open up  a new field  of research in the physics of nuclei 
with a  strangeness content \cite{suz04}.  The  state found in  the  
$K^{-}pnn$  system is  bound by 195 MeV.  It lives relatively long and 
 decays into a  hyperon and  two nucleons.  
Nuclear states of kaons were expected, since  early  studies of  
kaonic atoms    indicated that the nuclear potential for $K^{-}$ 
mesons is attractive, see review \cite{bat97}. 
On the other hand,  the nuclear absorption  related to the ($K,\pi$)
conversion was found  to be strong and  such   states  
would not be detectable  due to their  large ($\approx 100$ MeV) widths.  
The existence of detectable   narrow states was  thus limited to some 
high angular momentum states. Another possibility was indicated in 
Ref.~\cite{wyc86}. There it was shown that  $K^{-}$  mesons may be 
bound so strongly  that most of the 
decay channels would be 
 blocked. The mechanism of attraction was attributed to the 
$\Lambda(1405)$ state and  narrow states of about $100$ MeV binding were 
predicted to exist in large nuclei. That  study  was  motivated by the  
measurement of an  unusually strong  repulsive level shift found in the
2P states of $K^{-}$He  atoms \cite{bai83}.  
Recently, the existence  of narrow states   was predicted, by Akaishi and 
Yamazaki,  also in light 
nuclei such as $^4$He,  \cite{aka02},  and this prediction led to the 
experiment of 
Ref.~\cite{suz04}. Again,  the mechanism of attraction was attributed to 
the strong isospin $0$ attraction in the $K^{-}p$ channel, which 
generates the    $\Lambda(1405)$ as a  $K^{-}p$ bound state.

The actual experiment \cite{suz04}  shows that the   attraction due 
to the   $\Lambda(1405)$ apparently plays  an important role but is not 
strong enough to generate the 
binding as observed.  Both predictions  \cite{wyc86} and  \cite{aka02}   
are based on a  sizeable  proton component  in the nucleus and that   is not 
the case  in the  $K^{-}pnn$ system.
In this letter it is shown that two resonant states 
 $\Lambda(1405)$ and  $ \Sigma(1385)$ 
 coupled to  the $KN$ system may generate the  attraction 
required to produce the strong binding. 
This happens under two conditions: 

$\bullet$ 
the   $\Lambda(1405)$ and $ \Sigma(1385)$ in a nuclear medium  are 
located above  the $KN$ threshold. 

$\bullet$ 
 the binding of $K^{-}$ is    strong enough to block the main  
$\pi \Sigma$ and $ \pi \Lambda $ 
decay modes. 

\noindent Such  conditions may be  fulfilled in many nuclei, but 
the state generated with this interaction  will be  similar everywhere. 
The  meson is bound in a restricted small area around the nuclear centre.

\section{The origin  of $K^{-}$  attraction to  nuclei }

To present the argument,  let us assume   the  $K N  $  $S$-wave scattering 
amplitude to  be described in terms of a simple  resonance  formula 
\begin{equation}
\label{1} 
f_o =\frac{\gamma_o^2 }{E_{KN} - E_o + i\Gamma_o / 2 },
\end{equation}
where  $E_{KN}$
is the energy in the $ K N  $ channel and    $E_o$ is a  resonance 
energy. 
This amplitude is normalised to the scattering length at the threshold.
For kaons in  a  nuclear medium it  generates an optical potential 
\begin{equation}
\label{2} 
V_o(r)  =  \frac{4\pi}{2\mu_{KN}}  \rho(r) f_o,
\end{equation}
where  $\mu_{KN}$ is the $KN$  reduced mass and $\rho(r)$ is the nuclear density.
The resonant situation produces an attractive potential when $E_{KN} - E_o <0$ 
and a  repulsive one otherwise.  The question to settle is the value of  the 
actual energy of the resonance in the nuclear medium. This is  well 
understood in the $\Lambda(1405)$ case from the early days of $K^{-} $ atom studies. 
The state is located below the $K N  $ 
threshold of $1432$ MeV and leads to a repulsive  $K^{-}p$ scattering 
length. Contrary to
that, an attraction is  observed in kaonic atoms. It  is generated by two 
effects: 
(1) the energy of the $ KN$  system is reduced by the nuclear binding and the 
$ KN$  recoil with respect to the nucleus \cite{bar72};
 (2) the position of the resonance is shifted upwards as a result of  
Pauli blocking, \cite{alb76},\cite{wyc71},\cite{sta87}.  
Both these effects create  the  $E_{KN} - E_o <0 $ situation which results 
in the attractive potential.  
While the kaonic atoms involve the low density nuclear surface region, for  
deep binding one needs the  central densities. Calculations 
in Ref.\cite{wyc86} 
indicate  that in nuclear matter the upward shift of $\Lambda(1405)$ creates 
a similar situation of an  effective attraction. 
In the actual  calculations Eq.(\ref{1}) should be treated with more care as  
  $\Lambda(1405)$ is to be   described by a  two channel  
$K$-matrix which generates it as a $ KN$ quasi-bound state
\cite{dal80},\cite{mar81}. That results in a definite   energy dependence of
  $ \Gamma(E), E_o(E), \gamma(E)$, but the related physical effect 
remains the same.  
 
The isospin structure of the kaonic states  is  given by 
\begin{equation}
\label{3} 
\mid  K^{-} n > = \mid I = 1 >  , ~   
\mid  K^{-} p > =\frac{1}{2} \mid I = 1 >+ \frac{1}{2} \mid I = 0 > 
\end{equation}
and the  $I=1$ channel involves another resonance strongly  coupled to 
the $ KN$  channel in the subthreshold region. 
That is the  $J= 3/2 $ , $ \Sigma(1385)$    which involves   
 $K^{-}$ interactions with both  protons and  neutrons. In 
atomic  states  that resonance seems to play no role. However, in  deeply 
bound states it becomes the dominant factor. The  scattering 
amplitude 
\begin{equation}
\label{4} 
f_{  \Sigma } = 
2 {\bf p} {\bf p'}  \frac{\gamma_{\Sigma KN}^2 }
{E_{KN} - E_{\Sigma} + i\Gamma_{ \Sigma } / 2 } \ ,
\end{equation}
where  ${\bf p},  {\bf p'} $
are the relative momenta before and after the collisions,  is normalised to 
the standard $l+1/2$  partial wave 
which generates the factor 2. The resonant fraction gives 
 the scattering  volume.  Inside  the nuclear medium this 
term produces a dramatic effect on the kinetic energy of the meson. 
The disperison law in the  nuclear  medium  becomes 
\begin{equation}
\label{5} 
E_{K}^2 -m_{K}^2  = p^2 [ 1 + U_{ \Sigma }(E_{KN})] +  U_o \ , 
\end{equation}
where  $U_o= 2m_{K}V_o $ is the $S$-wave contribution to the potential, 
$E_{K} = m_{K}- E_{B}$ and 
\begin{equation}
\label{6} 
U_{  \Sigma }(E_{KN})  =  \frac{4 \pi\rho m_{K}  }{\mu_{KN}} 
\frac{2 \gamma_{\Sigma KN}^2(p^2)}{E_{KN} -E_{\Sigma } 
+ i\Gamma_{\Sigma }(E_{KN}) / 2 }
\end{equation}
describes the $P$-wave interaction due to the $ \Sigma(1385)$. 
This potential for energies $E_{KN}$ close to, but less than,  the resonance energy 
may be very large and attractive. Indeed,  so large as to dominate 
the kinetic energy term and make negative the $ p^2 $ term in the 
dispersion formula. That happens for  energy separations 
$E_{KN} -E_{\Sigma }$  of interest, at central nuclear densities and   
the coupling parameter which we discuss below.

The $ P$-wave formula for $ \Sigma(1385)$ coupling to the $KN$  channel 
requires more subtle control over the energy dependence of the width and the 
formfactor in the resonance formula (\ref{6}).
The width is composed of three terms  
\begin{equation}
\label{7} 
 \Gamma_{\Sigma} /2 = \Sigma_i ~p_i^3  \gamma_i^2(p_i^2) \ , 
\end{equation}
where the sum extends over the channels : $\pi \Sigma, \pi \Lambda, KN $. 
The resonant shape is  observed in the $ \pi \Lambda $ channel 
and the experimental width of 36 MeV gives the $\Sigma(1385)  \Lambda \pi $
coupling constant,  as the   decays to  
the  $\pi \Sigma $ channel constitute  only 13$\%$ of the total, 
Ref.\cite{PDG00}. Below the $ KN$ threshold there is no contribution  to the 
width coming from  this channel. However, Eq.~(\ref{7}) generates a definite 
contribution 
$  \mid p_{KN}\mid^3  \gamma_{KN}^2$ to the  position of the resonance 
$E_{\Sigma}$. This contribution  increases for energies 
below the resonance, making the  real part of 
\mbox{$ E_{KN} -E_{\Sigma } + i\Gamma_{\Sigma }(E) / 2 $}  
stay small  over a  considerable  range  of  energies.
The coupling parameter $ \gamma_{\Sigma \pi \Lambda}$   
is known from the resonance width  
$ \gamma_{\Sigma \pi \Lambda} ^2 = \Gamma_{\pi \Lambda}/ (2p_ {\pi \Lambda}^3)$
and the corresponding coupling to the  $ KN$ channel are related by
 SU(3) as 
$ \gamma_{\Sigma KN} ^2 / \gamma_{\Sigma \pi \Lambda} ^2 = 2/3$. 
The  experimental ratio of  0.51$\pm$0.18   obtained from $ K^{-}$ 
capture in deuterium \cite{bra77} is consistent with this  SU(3) value. 
These  numbers  indicate $( 1 + U_{  \Sigma } )$ to be negative at central 
nuclear densities for $E_{KN} < E_{\Sigma }$ and  
$\mid E_{KN} -E_{\Sigma } \mid  <  140 $ MeV. 
As shown in the  next 
section it is this effect that may generate the strong binding 
of $K^{-}$ mesons. In the nuclear 
matter case it  may generate  a dangerous almost-collapse mechanism. 
The safe-guard 
against that situation is the resonance denominator itself. Large 
binding generates large $E_{KN} -E_{\Sigma } $ in Eq.~(\ref{6}) 
i.e. a weaker  attraction 
 and at the end a finite saturation is 
obtained. The actual saturation point depends on the position of 
the $\Sigma $ resonance 
in the nuclear medium. As we show below it is likely to stay close to its
free  value.

The important point in question is the relative position of  the $KN$ 
threshold 
and the two resonances  $\Lambda(1405)$ and $ \Sigma(1385)$ situated in
nuclear matter. Calculations \cite{alb76},\cite{wyc71} produce an upward shift of
the $\Lambda(1405)$. 
This  shift  can  also be  extracted from  the experimental ratios of 
 $\pi^+ \Sigma^-$/  $ \pi^-  \Sigma^+$  pairs produced by kaons stopped 
in  nuclear emulsion. Such an analysis  indicates  
a shift  of about 10 MeV at the 
extreme nuclear
surface\cite{sta87}.   Here we use an alternative,  but fully consistent
estimate.  The coupling $G_{\Lambda KN} \Psi_{N} \Psi_{ \Lambda} \psi_K$  
may be described
by a constant   $G_{\Sigma KN}^2/4\pi \approx 1$ \cite{dum83}.  This generates 
a $K$-exchange  potential between a proton and  $\Lambda(1405)$ of an inverse range 
$\kappa=\sqrt{ ( M_{\Lambda}-M_K )^2 -m_K^2} $ and an average nuclear potential 
for  the  $\Lambda(1405)$ of 
 strength $ V_{\Lambda} = \rho_p G^2 /\kappa^2 \approx 200 MeV $ 
at the nuclear centres. 

The  shift of the  $\Sigma(1385)$ follows from the gradient coupling 
$ G_{\Sigma KN}\Psi_{N} \Psi_{ \Sigma}^{\mu} \nabla_{\mu}\psi_K $. Such 
a  coupling generates $K$-exchange between a  nucleon  and the $ \Sigma$ 
which 
flips  the spin of the hyperon  but also generates a diagonal term. The latter 
  requires a correction at
short nucleon $ \Sigma$  distances which is to be introduced in a  way 
analogous to the 
subtraction of $\delta(r)$ in the $\pi$-exchange  contribution to the 
nucleon-nucleon potential. The net potential, 
averaged over the  nucleon densities, generates an attractive  nuclear  
well for the 
$ \Sigma(1385)$  with  the strength 
$V_{\Sigma}= - \rho_{1}  G_{\Sigma KN}^2/3 $. Here 
the density $\rho_{1}  $ involves that component of the nuclear matter 
which couples  
to the hyperon in the isospin 1 state that is mostly neutrons. This  potential
is rather weak being about  --30 MeV depth in nuclear matter. 
There is, unfortunately, no direct  experimental check on this resonant shift.

\section{The  $K^{-}NNN $ system } 

In the $K^{-}pnn $ situation the $\Lambda(1405)$ may be formed on the proton. 
The average nuclear potential for the resonance vanishes within the 
$K$-meson  
exchange $\Lambda(1405)$-$p$ interaction  of the previous section. 
The  $ \Sigma(1385)$ is formed 
with much  higher probability since, as seen from Eq.(\ref{3}), it involves
mostly   neutrons.  With the  $K$-meson-exchange model   one 
finds the nuclear well for the $ \Sigma(1385)$ to be about $- 20$  MeV deep.  

With the  $K^{-}$  bound by some  200 MeV the relative separation of the 
$ KN$ threshold and the  $\Lambda(1405)$ position amounts  to 170 MeV and 
an   extrapolation of  the scattering amplitudes to this region 
is necessary. The $K$-matrix parametrization  of A.Martin\cite{mar81}
( $K_{NN}= -1.65 fm,  R_{NN}= 0.18 fm$ ) is used here,  
but it is supplemented by a separable model of Ref.\cite{krz75}  to obtain 
a smooth subthreshold extrapolation. This  procedure generates the 
$I=0$ scattering amplitude of  $f_0 = 1.48 $ fm, a fairly standard result in 
this energy  region. As the  energy of $ KN$ is so low,  the pionic decay
channels are closed and the scattering amplitudes are real. 
For the $I=1$,  $K^{-}n $  amplitude the solution also from  Ref.~\cite{mar81}
is used 
($ K_{NN}= 1.07$ fm) which produces $f_1 = 0.34 $ fm. These values 
generate a $V_o $, the  potential well for $K^{-}$,    of 
 $ -110$ MeV depth at the centre  of the 
tritium  nucleus. This is far too weak to produce  the experimental binding.  
An additional and, in fact, predominant  attraction comes from the 
$ \Sigma(1385)$.
It generates the potential  described  by a gradient term
\begin{equation}
\label{7a} 
 U_{G}  =  \stackrel{\leftarrow}{\nabla} U_{ \Sigma }(E_{KN})
\stackrel{\rightarrow}{\nabla} \ , 
\end{equation}
where $ U_{  \Sigma } $ is given by the resonant 
amplitude of Eq.~(\ref{6}). 

Now we  look for the variational solution for the kaonic  energy level,  
$ \epsilon = E_K ^2 - m_K^2, $
 
\begin{equation}
\label{8} 
\epsilon  = Min   \int d{\bf r} \Psi(r)
[  p^2  + 
 \stackrel{\leftarrow}{\nabla} U_{ \Sigma }(E_{KN})
\stackrel{\rightarrow}{\nabla}   +  U_o]  \Psi(r). 
\end{equation}  
In order to understand the nature of the  solution and to guess the 
proper trial  wave function we first solve a simpler problem. 
Consider  a square well of radius $R$  with the potential given 
by our local and gradient terms. In the internal region the  wave 
equation  
\begin{equation}
\label{9} 
-[1 +  U_{ \Sigma }(E_{KN})] \psi''_i(r)   +  U_o \psi_i(r)
= \epsilon \psi_i(r)
\end{equation}
may,  for some energies,  be characterised by negative values of 
$[1 +  U_{ \Sigma }(E_{KN})]$. One has $\psi_i = \Psi/r = \sin(\kappa r) $  with 
$ \kappa^2 = (-\epsilon + U_o )/(1 +  U_{ \Sigma }(E_{KN})) $.
In the external  region 
\begin{equation}
\label{9a} 
-\psi''_e(r) = \epsilon \psi_e(r)
\end{equation}
gives the asymptotic  solution $\psi_e = C \exp(-\sqrt{-\epsilon}r) $, 
which has to be continuously matched to the internal solution 
that gives  the eigenvalue condition 
$ \kappa cot(\kappa R) = - \sqrt{-\epsilon} $. It leads  to the solution 
\begin{equation}
\label{10} 
\epsilon =  U_o +  [1 +  U_{ \Sigma }(E_{KN})] ( \frac {\pi \xi} {R})^2 \ ,  
\end{equation}
where $\xi$  is a number in the  range [0.5,1].
For   negative and  energy independent $1 +  U_{ \Sigma }$ the system 
collapses,  since the minimal ( and infinite)  energy is obtained 
in the limit 
$R \rightarrow 0$. That is not the case with the  realistic 
$U_{ \Sigma }(E)$.  Eq.~(\ref{10})  presents a nonlinear 
problem with respect to $\epsilon$ since  
$ E_{KN} = M_N + m_K - E_B - E_{BN} $, 
where $E_B, E_{BN}$ are the $K^{-} $ and $N$  binding energies. 
The minimum is found numerically and the solutions are 
always finite. With the parameters 
given in the text one obtains a trajectory for  the binding  energy 
  $ E_B(R) =  -200$ MeV + $ R~  27$ MeV/fm,
which is valid for  radii $R$ characteristic in tritium. 

The minimal energy is obtained  with $ R \rightarrow 0$. 
This  limit  changes,  when  a formfactor is introduced in the coupling 
parameter $ \gamma_{\Sigma KN}(p^2) $ of Eq.~(\ref{6}).

It is interesting to compare the $K^{-}pnn $
system discussed above with the $K^{-}ppn $ one. In the latter case,   
the central attraction is stronger due to the two protons that 
may generate the  
$\Lambda(1405)$. One obtains now  $ V_o(0)= -140$ MeV.  On the other hand 
the $ \Sigma(1385)$  gradient term is  weaker by 25$\%$ due to the  smaller 
$I=1$ content, as given by Eq.(\ref{3}).  The $K^{-} $ binding   
trajectory becomes   $ E_B(R) = -173$ MeV +$R~ 10$ MeV /fm. 
The difference in the binding of these two systems may thus determine 
$R$ that is the size of the region to which the $K^{-} $ meson is confined. 
  
Now  the  variational wave function is used  in the form 
given by the square well model $\Psi = \psi/r$, and  the 
free  parameter of the variational procedure is $R$. 
The procedure itself  is as follows:

\noindent1)  For a given radius $R$ the wave function $\Psi$ is obtained.  

\noindent2) The expected value of $\epsilon $ in Eq.~(\ref{8}) is calculated. 

\noindent3) The kaon binding energy $ E_B$ is varied in the 
potential term $  U_{ \Sigma }(E_{KN}) $ until  selfconsistency 
with the variational binding is achieved.  

\noindent4) The  minimum of $ E_B$ with respect to $R$ is found.

The radii $R$ in the small region $ R <0.5 $ fm  generate  essentially 
equivalent binding energies.  
A  formfactor for $\Sigma(1385)$,   
$ \gamma^{2}_{\Sigma KN}(p^2) =  \gamma^{2}_{\Sigma KN}(0)
/( 1+ (pr_o)^2)$
introduced into Eq.~(\ref{6}), 
locates  the minimum at finite values of $R$.  With the  characteristic 
momenta of the confined kaons $ p \sim  1/R$ and $ r_o \sim 0.5$ fm 
one obtains a very shallow minimum of $ E_B(R)$ at $ R = 0.4$ fm. 
The results are given in Table I.
These include the $K$ meson binding and a $7$ MeV binding due to nucleons. 
The calculated energy is somewhat smaller than the 
experimental one. The difference may be due to many uncertainties in 
the physics involved. These are: 

\noindent1) An uncertain subthreshold extrapolation of the resonant $ I=0$ and 
nonresonant $I=1$ scattering amplitudes. 

\noindent2) No experimental control   of the  $ \Sigma(1385)$  energy in the nucleus. 

\noindent3) An uncertain $  \gamma_{\Sigma KN}$ coupling (an effect of this 
uncertainty is indicated in Table I). 

\noindent4) Some restructuring of the three nucleon state. 

Finally,  more parameters in the trial wave function may reduce the 
variational result.  We are not attempting any further discussion 
of these uncertainties.  The aim of this letter is to show that 
the  two $KN$ subthreshold resonances are sufficient to bind 
the $KNNN$ system as strongly as observed.   
The  width of  such a state  predicted in   
Ref.\cite{wyc86} is consistent with the experimental finding. 
It is based on the  extrapolation (by the phase space)  
of the $KNN \rightarrow  \Sigma(\Lambda) N$ decay rate  
known from the nuclear emulsion studies.

\begin{table}
\caption{The binding energies $E_B$, in MeV, of the $K^{-}pnn$  and $K^{-}ppn $ systems.
First line is based on the $ SU(3)$ value for the $KN  \Sigma(1385)$  
coupling.  In the second line this coupling is 
enhanced by $20\%$. }
\begin{tabular}{ccc}
$\gamma_{\Sigma KN} ^2 / \gamma_{\Sigma \pi \Lambda} ^2$  & $K^{-}pnn $  &  $K^{-}ppn $   \\
2/3    & 169   &  155       \\
1.2*2/3    & 192   &  175      \\
\end{tabular}
\label{table1}
\end{table}

\section{Conclusions} 

To summarise let us indicate again the main result, and suggest  some 
topics for  further research. 

$\bullet$ The 
strange  $\Lambda(1405)$ and $ \Sigma(1385)$ baryonic    
states  coupled to  the $KN$ system may generate  
the  strongly bound $K^{-} $ states, as is  observed. 
Such $K^{-} $  states, under normal nuclear density tend  to be localised 
close to the nuclear centres. For much higher densities these are not
necessarily the best  solutions and the attraction due to  the 
 $ \Sigma(1385)$ state may indicate a  proximity to kaon condensation. 

$\bullet$ The model developed here requires confirmation of some 
parameters, in  particular the validity of the SU(3) symmetry in 
the  $ \Sigma(1385)$  couplings 
and the position of this resonance in  nuclear media. 
An experiment indicating the $ \Sigma(1385)$ decays in nuclei might 
elucidate this question.

$\bullet$ This model stresses the strengths  of  
the $K^{-}n $ interaction. 
The search for $K^{-}nn $, $K^{-}nnn $ and other  objects of
 neutron excess could be helpful. 

One of the authors (S.W.) wishes to acknowledge the hospitality of the
Helsinki Institute of  Physics, where part of this
work was carried out. The authors also thank Dr. Ryugo Hayano 
for useful correspondence. This project is financed by the Academy of 
Finland contract 54038,  and the European Community Human 
potential Program HPRN-2002-00311 EURIDICE.

%\begin{references}

\end{document}